\documentclass[12pt]{article}
\usepackage{graphics,color}
\begin{document}
\thispagestyle{empty}
\noindent\
\\
\\
\\
\begin{center}
\large \bf Lepton Mixing and the Neutrino Mixing Angle $\theta^{}_{13}$
\end{center}
\hfill
 \vspace*{1cm}
\noindent
\begin{center}
{\bf Harald Fritzsch}\\
Department f\"ur Physik, Universit\"at M\"unchen,\\
Theresienstra{\ss}e 37, 80333 M\"unchen
\vspace*{0.5cm}
\end{center}

\begin{abstract}

We discuss the neutrino oscillations,
using texture zero mass matrices for the leptons,
including radiative correction. The
neutrino mixing angle $\theta^{}_{13}$
is calculated and agrees with the
result of the new Daya Bay experiment.
\end{abstract}

\newpage

Now it is known that the three neutrinos have finite masses.
The neutrinos, emitted in weak decays, are mixtures of
different mass eigenstates. This leads to leptonic flavor mixing, in particual to neutrino
oscillations.\\

The lepton flavor
mixing is described by a $3\times 3$ unitary matrix $U$, analogous to the CKM mixing matrix
for the quarks:\\
\begin{eqnarray}
-{\cal L}^{}_{\rm cc} = \frac{g}{\sqrt{2}} \ \overline{
\left(\matrix{e & \mu & \tau} \right)^{}_{\rm L}} \ \gamma^\mu
\left( \matrix{ U^{}_{e1} & U^{}_{e2} & U^{}_{e3} \cr U^{}_{\mu 1} &
U^{}_{\mu 2} & U^{}_{\mu 3} \cr U^{}_{\tau 1} & U^{}_{\tau 2} &
U^{}_{\tau 3} \cr} \right) \left(\matrix{ \nu^{}_1 \cr \nu^{}_2 \cr
\nu^{}_3 \cr} \right)^{}_{\rm L} W^-_\mu + {\rm h.c.} \; .
\end{eqnarray}
\\
It can be parametrized in terms of three angles and three phases:\\
\begin{eqnarray}
U = \left( \matrix{ c^{}_{12}
c^{}_{13} & s^{}_{12} c^{}_{13} & s^{}_{13} e^{-i\delta} \cr
-s^{}_{12} c^{}_{23} - c^{}_{12} s^{}_{13} s^{}_{23} e^{i\delta} &
c^{}_{12} c^{}_{23} - s^{}_{12} s^{}_{13} s^{}_{23} e^{i\delta} &
c^{}_{13} s^{}_{23} \cr s^{}_{12} s^{}_{23} - c^{}_{12} s^{}_{13}
c^{}_{23} e^{i\delta} & -c^{}_{12} s^{}_{23} - s^{}_{12} s^{}_{13}
c^{}_{23} e^{i\delta} & c^{}_{13} c^{}_{23} \cr} \right) P^{}_\nu \;
\end{eqnarray}
\\
where $c^{}_{ij} \equiv \cos\theta^{}_{ij}$, $s^{}_{ij} \equiv
\sin\theta^{}_{ij}$ (for $ij = 12, 13, 23$). The phase matrix $P^{}_\nu ={\rm
Diag}\{e^{i\rho}, e^{i\sigma}, 1\}$ is relevant only, if
the neutrino masses are Majorana masses.\\

The neutrino oscillations are described by two large mixing angles:
$\theta^{}_{12} \simeq 34^\circ$ and $\theta^{}_{23} \simeq
45^\circ$\ . The third mixing angle
$\theta^{}_{13}$ must be much smaller. The results of the experiments T2K (ref. (1)), MINOS (ref. (2)) and
Double Chooz (ref. (3)) indicate that this angle is about $8^\circ$. The three CP-violating phases are
unknown.\\

The Daya Bay Collaboration has recently measured the
neutrino mixing angle $\theta^{}_{13}$ (ref. 4):
\begin{eqnarray}
\sin^2 2\theta^{}_{13} = 0.092 \pm 0.016 ({\rm stat})
\pm 0.005 ({\rm syst}) \;,
\end{eqnarray}
which gives $\theta^{}_{13} \simeq 8.8^\circ \pm
0.8^\circ$.\\

In this letter we use texture zero mass matrices and calculate the angle $\theta^{}_{13}$.
However we prefer another parametrization of the unitary mixing matrix (ref. (5)):\\

\begin{eqnarray}
U = \left ( \matrix{ s^{}_l s^{}_{\nu} c + c^{}_l c^{}_{\nu}
e^{-i\varphi} & s^{}_l c^{}_{\nu} c - c^{}_l s^{}_{\nu}
e^{-i\varphi} & s^{}_l s \cr c^{}_l s^{}_{\nu} c - s^{}_l c^{}_{\nu}
e^{-i\varphi} & c^{}_l c^{}_{\nu} c + s^{}_l s^{}_{\nu}
e^{-i\varphi} & c^{}_l s \cr - s^{}_{\nu} s   & - c^{}_{\nu} s   & c
\cr } \right ) P^{}_\nu \; ,
\end{eqnarray}
\\
where $c^{}_{l,\nu} \equiv \cos\theta^{}_{l,\nu}$, $s^{~}_{l,\nu}
\equiv \sin\theta^{}_{l,\nu}$, $c \equiv \cos\theta$ and $s \equiv
\sin\theta$. The angle $\theta^{}_{\nu}$ is the solar angle $\theta^{}_{sun}$, the angle
$\theta$ is the atmospheric angle $\theta^{}_{at}$, and the angle
$\theta^{}_{l}$ is the "reactor angle", measured in the Daya Bay experiment.\\

We assume that the mass matrices of the leptons and quarks are texture zero matrices: \\

\begin{eqnarray}
M= \left( \matrix{ 0 & A & 0 \cr A^* & B & C \cr
0 & C^* & D \cr} \right) \;.
\end{eqnarray}
\\

We find for the leptonic mixing angles (ref. (6)):\\
\\

\begin{equation}
\tan\theta^{}_l \simeq
\sqrt{m^{}_e/m^{}_\mu}
\end{equation}
\begin{equation}
\tan\theta^{}_\nu \simeq
\sqrt{m^{}_1/m^{}_2}.
\end{equation}
\\

The solar angle $\theta^{}_{sun}$ is measured to about $33^\circ$:\\

\begin{equation}
\tan33^\circ\simeq
\sqrt{m^{}_1/m^{}_2}.
\end{equation}
\\

Thus we obtain for the neutrino mass ratio: \\

\begin{equation}
{m^{}_1/m^{}_2} \simeq 0.42.
\end{equation}
\\
This relation and the experimental results for the mass differences of the neutrinos, measured
by the neutrino oscillations, allow us to
determine the three neutrino masses  ( ref. (7)):\\
\begin{eqnarray}
{m^{}_1} \simeq 0.004~eV , \nonumber \\
{m^{}_2} \simeq 0.01~eV , \nonumber \\
{m^{}_3} \simeq 0.05~eV.
\end{eqnarray}
\\

We can also calculate the mixing angle $\theta^{}_{13}$, using eq.
(4). It is about $2.8^\circ$ ( see ref. (7)). The angle, measured in
the Daya Bay experiment, is about a factor three
larger.\\

The texture zero mass matrices, given in eq. (5), describe very well the flavor mixing of quarks ( ref. (8)).
However we expect that the mass matrices of the quarks and leptons are not exactly given by
texture zero matrices. Higher order contributions of the
order of the fine-structure constant $\alpha$ will contribute, in particular the zeros will be replaced by small numbers.\\

In a good approximation the ratios of the masses of the quarks with the same electric charge
are universal:
\\

 \begin{equation}
\frac{m^{}_u}{m^{}_c} \simeq \frac{m^{}_c}{m^{}_t} \simeq 0.005,
\end{equation}
\begin{equation}
\frac{m^{}_d}{m^{}_s} \simeq \frac{m^{}_s}{m^{}_b} \simeq 0.044.
\end{equation}
\\

The dynamical reason for this universality is unclear. It might follow
from specific properties of the texture zero mass matrices.  But in the
case of the leptons there is no universality:\\

 \begin{equation}
\frac{m^{}_e}{m^{}_\mu} \simeq 0.005,
\end{equation}
 \begin{equation}
\frac{m^{}_\mu}{m^{}_\tau} \simeq 0.06.
\end{equation}
\\

If the ratios of the lepton masses were universal, the mass of the electron would have
to be twelve times larger than observed.\\

As mentioned above, the universality is not expected to
be exact, due to radiative corrections, related to the electroweak theory.
A radiative correction of the order of $(\alpha/\pi){m^{}_\tau}\simeq 4~MeV $ would have
to be added to the charged
lepton masses. Such a contribution is relatively small for the muon and the tauon, but large
in comparison to the observed electron mass. One expects that the physical electron mass
is the sum of a bare electron mass, due to the texture zero mass matrix, and a radiative
correction, which could either be positive or negative:\\

\begin{equation}
{m^{}_e} = B - R
\end{equation}
\\

Here B is the bare electron mass and R the radiative correction. An example would
be: B= 5.3 MeV, R=4.8 MeV.  Both terms nearly cancel each other - the physical electron
mass is very small.\\

Radiative corrections also contribute to the muon and the tauon mass, but
here the corrections are small in comparison to the bare masses. We shall use for the bare muon mass
the same value of R as for the electron:\\

$M^{}_\mu$ = $m^{}_\mu + R\simeq 110.5~MeV$. \\

We calculate the angle $\theta^{}_{13}$ ( see. eq. (4)):\\

\begin{equation}
\tan\theta^{}_l \simeq
\sqrt{M^{}_e/M^{}_\mu}\simeq\sqrt{5.3/100.9}\simeq 0.22
\end{equation}

\begin{equation}
\theta^{}_{13}\simeq 8.9^\circ
\end{equation}
\\
This angle agrees with the result of the Daya Bay experiment.\\

The texture zero mass matrices seem
to work well not only for the quarks, but also for the leptons. The neutrino masses
are less hierarchical than the charged lepton masses, and this leads to the large
mixing angles, observed in the neutrino oscillations. The ratio of the electron mass and the
muon mass is very small, and this leads to the small angle $\theta^{}_{13}$.\\

Acknowledgement: I thank Prof. K. K. Phua for the support during my visit of the
Institute of Advanced Study at the Nanyang Technological University in Singapore. \\

\end{document}